\documentclass[aps,reprint,superscriptaddress,nofootinbib,amsmath,amssymb,floatfix]{revtex4-2}
\usepackage{graphicx}     
\usepackage{dcolumn}      
\usepackage{bm}           
\usepackage{hyperref}     
\usepackage{xcolor}       
\usepackage{upgreek}      
\usepackage{xspace}
\usepackage{comment}      

\usepackage{amsmath}
\usepackage{mathtools}

\usepackage{lineno}       
\modulolinenumbers[1]     

\hypersetup{
    colorlinks=true,
    linkcolor=blue,
    citecolor=blue,
    urlcolor=blue
}

\begin{document}
%

\newcommand{\pp}           {pp\xspace}
\newcommand{\ppbar}        {\mbox{$\mathrm {p\overline{p}}$}\xspace}
\newcommand{\XeXe}         {\mbox{Xe--Xe}\xspace}
\newcommand{\PbPb}         {\mbox{Pb--Pb}\xspace}
\newcommand{\pA}           {\mbox{pA}\xspace}
\newcommand{\pPb}          {\mbox{p--Pb}\xspace}
\newcommand{\AuAu}         {\mbox{Au--Au}\xspace}
\newcommand{\dAu}          {\mbox{d--Au}\xspace}

\newcommand{\s}            {\ensuremath{\sqrt{s}}\xspace}
\newcommand{\snn}          {\ensuremath{\sqrt{s_{\mathrm{NN}}}}\xspace}
\newcommand{\pt}           {\ensuremath{p_{\rm T}}\xspace}
\newcommand{\meanpt}       {$\langle p_{\mathrm{T}}\rangle$\xspace}
\newcommand{\ycms}         {\ensuremath{y_{\rm CMS}}\xspace}
\newcommand{\ylab}         {\ensuremath{y_{\rm lab}}\xspace}
\newcommand{\etarange}[1]  {\mbox{$\left | \eta \right |~<~#1$}}
\newcommand{\yrange}[1]    {\mbox{$\left | y \right |~<~#1$}}
\newcommand{\dndy}         {\ensuremath{\mathrm{d}N_\mathrm{ch}/\mathrm{d}y}\xspace}
\newcommand{\dndeta}       {\ensuremath{\mathrm{d}N_\mathrm{ch}/\mathrm{d}\eta}\xspace}
\newcommand{\lowmult}      {\ensuremath{\langle dN_{\text{ch}}/d\eta \rangle \lesssim 5}\xspace}
\newcommand{\avdndeta}     {\ensuremath{\langle\dndeta\rangle}\xspace}
\newcommand{\dNdy}         {\ensuremath{\mathrm{d}N_\mathrm{ch}/\mathrm{d}y}\xspace}
\newcommand{\Npart}        {\ensuremath{N_\mathrm{part}}\xspace}
\newcommand{\Ncoll}        {\ensuremath{N_\mathrm{coll}}\xspace}
\newcommand{\dEdx}         {\ensuremath{\textrm{d}E/\textrm{d}x}\xspace}
\newcommand{\RpPb}         {\ensuremath{R_{\rm pPb}}\xspace}
\newcommand{\Minv}         {\ensuremath{M_{\uppi^{+}\uppi^{-}}}\xspace}
\newcommand{\ctaueff}{\mbox{$\left\langle c\tau_{\mathrm{res}}^{\mathrm{eff}} \right\rangle$}\xspace}
\newcommand{\masseff}{\mbox{$\left\langle m_{\mathrm{res}}^{\mathrm{eff}} \right\rangle$}\xspace}

\newcommand{\nineH}           {$\sqrt{s}~=~0.9$~Te\kern-.1emV\xspace}
\newcommand{\seven}           {$\sqrt{s}~=~7$~Te\kern-.1emV\xspace}
\newcommand{\eight}            {$\sqrt{s}~=~8$~Te\kern-.1emV\xspace}
\newcommand{\thirteen}        {$\sqrt{s}~=~13$~Te\kern-.1emV\xspace}
\newcommand{\thirteensix}    {$\sqrt{s}~=~13.6$~Te\kern-.1emV\xspace}
\newcommand{\fourteen}        {$\sqrt{s}~=~14$~Te\kern-.1emV\xspace}
\newcommand{\twoH}            {$\sqrt{s}~=~0.2$~Te\kern-.1emV\xspace}
\newcommand{\twosevensix}  {$\sqrt{s}~=~2.76$~Te\kern-.1emV\xspace}
\newcommand{\five}               {$\sqrt{s}~=~5.02$~Te\kern-.1emV\xspace}
\newcommand{\fivethirtysix}   {$\sqrt{s}~=~5.36$~Te\kern-.1emV\xspace}
\newcommand{\fivefive}         {$\sqrt{s}~=~5.5$~Te\kern-.1emV\xspace}
\newcommand{\twosevensixnn}{$\sqrt{s_{\mathrm{NN}}}~=~2.76$~Te\kern-.1emV\xspace}
\newcommand{\fivenn}           {$\sqrt{s_{\mathrm{NN}}}~=~5.02$~Te\kern-.1emV\xspace}
\newcommand{\fivethirtysixnn}   {$\sqrt{s_{\mathrm{NN}}}~=~5.36$~Te\kern-.1emV\xspace}
\newcommand{\fivefivenn}         {$\sqrt{s_{\mathrm{NN}}}~=~5.5$~Te\kern-.1emV\xspace}
\newcommand{\LT}           {L{\'e}vy-Tsallis\xspace}
\newcommand{\GeVc}         {Ge\kern-.1emV/$c$\xspace}
\newcommand{\MeVc}         {Me\kern-.1emV/$c$\xspace}
\newcommand{\TeV}          {Te\kern-.1emV\xspace}
\newcommand{\GeV}          {Ge\kern-.1emV\xspace}
\newcommand{\MeV}          {Me\kern-.1emV\xspace}
\newcommand{\tev}          {Te\kern-.1emV\xspace}
\newcommand{\gev}          {\rm Ge\kern-.1emV\xspace}
\newcommand{\mev}          {\rm Me\kern-.1emV\xspace}
\newcommand{\GeVmass}      {Ge\kern-.2emV/$c^2$\xspace}
\newcommand{\MeVmass}      {Me\kern-.2emV/$c^2$\xspace}
\newcommand{\lumi}         {\ensuremath{\mathcal{L}}\xspace}

\newcommand{\ITS}          {\rm{ITS}\xspace}
\newcommand{\TOF}          {\rm{TOF}\xspace}
\newcommand{\ZDC}          {\rm{ZDC}\xspace}
\newcommand{\ZDCs}         {\rm{ZDCs}\xspace}
\newcommand{\ZNA}          {\rm{ZNA}\xspace}
\newcommand{\ZNC}          {\rm{ZNC}\xspace}
\newcommand{\SPD}          {\rm{SPD}\xspace}
\newcommand{\SDD}          {\rm{SDD}\xspace}
\newcommand{\SSD}          {\rm{SSD}\xspace}
\newcommand{\TPC}          {\rm{TPC}\xspace}
\newcommand{\TRD}          {\rm{TRD}\xspace}
\newcommand{\VZERO}        {\rm{V0}\xspace}
\newcommand{\VZEROA}       {\rm{V0A}\xspace}
\newcommand{\VZEROC}       {\rm{V0C}\xspace}
\newcommand{\Vdecay} 	   {\ensuremath{V^{0}}\xspace}

\newcommand{\ee}           {\ensuremath{e^{+}e^{-}}} 
\newcommand{\pip}          {\ensuremath{\uppi^{+}}\xspace}
\newcommand{\pim}          {\ensuremath{\uppi^{-}}\xspace}
\newcommand{\Cpi}          {\ensuremath{\uppi^{\pm}\xspace}}
\newcommand{\kap}          {\ensuremath{\rm{K}^{+}}\xspace}
\newcommand{\kam}          {\ensuremath{\rm{K}^{-}}\xspace}
\newcommand{\pbar}         {\ensuremath{\rm\overline{p}}\xspace}
\newcommand{\kzero}        {\ensuremath{{\rm K}^{0}_{\rm{S}}}\xspace}
\newcommand{\lmb}          {\ensuremath{\Lambda}\xspace}
\newcommand{\almb}         {\ensuremath{\overline{\Lambda}}\xspace}
\newcommand{\Om}           {\ensuremath{\Omega^-}\xspace}
\newcommand{\Mo}           {\ensuremath{\overline{\Omega}^+}\xspace}
\newcommand{\X}            {\ensuremath{\Xi^-}\xspace}
\newcommand{\Ix}           {\ensuremath{\overline{\Xi}^+}\xspace}
\newcommand{\Xis}          {\ensuremath{\Xi^{\pm}}\xspace}
\newcommand{\Oms}          {\ensuremath{\Omega^{\pm}}\xspace}
\newcommand{\degree}       {\ensuremath{^{\rm o}}\xspace}
\newcommand{\Dzero}     {\ensuremath{\rm D}^{0}\xspace}
\newcommand{\Dplus}     {\ensuremath{\rm D}^{+}\xspace}
\newcommand{\Dstar}     {\ensuremath{\rm D}^{*+}\xspace}
\newcommand{\Dstarwide} {\ensuremath{\rm D}^{*}(2010)^{+}\xspace}
\newcommand{\Lc}        {\ensuremath{\rm \Lambda}_{\rm c}^{+}\xspace}
\newcommand{\Sc}        {\ensuremath{\rm \Sigma}_{\rm c}^{0,++}\xspace}

\newcommand{\aprot}{\ensuremath{\mathrm{\bar{p}}}\,\xspace}
\newcommand{\prot}{\ensuremath{\mathrm{p}}\,\xspace}
\newcommand{\pP}{\ensuremath{\mathrm {p\mbox{--}p}}\,\xspace}
\newcommand{\pN}{\ensuremath{\mathrm {p\mbox{--}n}}\,\xspace}
\newcommand{\pXi}{\ensuremath{\mathrm {p\mbox{--}\Xi}}\,\xspace}
\newcommand{\pOmega}{\ensuremath{\mathrm {p\mbox{--}\Omega}}\,\xspace}
\newcommand{\LXi}{\ensuremath{\mathrm {\Lambda\mbox{--}\Xi}}\,\xspace}
\newcommand{\phiP}{\ensuremath{\mathrm {\upphi\mbox{--}p}}\,\xspace}
\newcommand{\akP}{\ensuremath{\mathrm {K^-\mbox{--}p}}\,\xspace}
\newcommand{\Kp}{\ensuremath{\mathrm {K\mbox{--}p}}\,\xspace}
\newcommand{\aKp}{\ensuremath{\mathrm {K^+\mbox{--}p}}\,\xspace}
\newcommand{\Kap}{\ensuremath{\mathrm {K^-\mbox{--}\overline{p}}}\,\xspace}
\newcommand{\pipi}{\ensuremath{\mathrm {\uppi^+\mbox{--}\uppi^+}}\,\xspace}
\newcommand{\SCpipi}{\ensuremath{\mathrm {\uppi^{\pm}\mbox{--}\uppi^{\pm}}}\,\xspace}
\newcommand{\apiapi}{\ensuremath{\mathrm {\uppi^-\mbox{--}\uppi^-}}\,\xspace}
\newcommand{\spipi}{\ensuremath{\mbox{$\uppi$--$\uppi$}}\,\xspace}
\newcommand{\pap}{\ensuremath{\mathrm {p\mbox{--}\bar{p}}}\,\xspace}
\newcommand{\paL}{\ensuremath{\mathrm {p\mbox{--}\overline{\Lambda}}}\,\xspace}
\newcommand{\apL}{\ensuremath{\mathrm {\overline{p}\mbox{--}\Lambda}}\,\xspace}
\newcommand{\LaL}{\ensuremath{\mathrm {\Lambda\mbox{--} \overline{\Lambda}}}\,\xspace}
\newcommand{\pL}{\ensuremath{\mathrm {p\mbox{--}\Lambda}}\,\xspace}
\newcommand{\ApaL}{\ensuremath{\mathrm {\bar{p}\mbox{--}\overline{\Lambda}}}\,\xspace}
\newcommand{\BBbar}{\ensuremath{\mathrm {B\mbox{--} \bar{B}}}\,\xspace}
\newcommand{\BB}{\ensuremath{\mathrm {B\mbox{--} B}}\,\xspace}
\newcommand{\rhop}{\ensuremath{\mathrm {\uprho^{0}\mbox{--} p}}\,\xspace}
\newcommand{\rhoap}{\ensuremath{\mathrm {\uprho^{0}\mbox{--} \overline{p}}}\,\xspace}

\newcommand{\rs}           {\ensuremath{r^*}\xspace}
\newcommand{\rsv}           {\ensuremath{\vec{r}^*}\xspace}
\newcommand{\rc}           {\ensuremath{r_\mathrm{core}}\xspace}
\newcommand{\rcv}           {\ensuremath{\vec{r}_\mathrm{core}}\xspace}
\newcommand{\rcs}           {\ensuremath{r^*_\mathrm{core}}\xspace}
\newcommand{\rcsv}           {\ensuremath{\vec{r}^*_\mathrm{core}}\xspace}
\newcommand{\ks}           {\ensuremath{k^*}\xspace}
\newcommand{\ksSq}           {\ensuremath{k^{*2}}\xspace}
\newcommand{\ksv}           {\ensuremath{\vec{k}^*}\xspace}
\newcommand{\Sr}           {\ensuremath{S(r)}\xspace}
\newcommand{\Ck}           {\ensuremath{C(k)}\xspace}
\newcommand{\Srs}           {\ensuremath{S(\rs)}\xspace}
\newcommand{\Cks}           {\ensuremath{C(\ks)}\xspace}
\newcommand{\mt}           {\ensuremath{m_{\mathrm{T}}}\xspace}


\title{Light Antinuclei Coalescence: Femtoscopic Constraints via Neural-Flow Surrogates}

\author{M. Korwieser}
\email{max.korwieser@tum.de} 
\affiliation{Department of Physics, Technical University of Munich, Garching at Munich, Germany}

\author{L. Fabbietti}
\affiliation{Department of Physics, Technical University of Munich, Garching at Munich, Germany}

\author{B. Hashemi}
\affiliation{Max Planck Institute for Mathematics in the Sciences, Leipzig, Germany}

\author{L. Heinrich}
\affiliation{Department of Physics, Technical University of Munich, Garching at Munich, Germany}

\author{M. Mahlein}
\affiliation{Department of Physics, Technical University of Munich, Garching at Munich, Germany}

\author{D. L. Mihaylov}
\affiliation{Faculty of Physics, Sofia University ``St. Kliment Ohridski'', Sofia, Bulgaria}

\author{C. S. Zeyn}
\affiliation{Department of Physics, Technical University of Munich, Garching at Munich, Germany}

\date{July 06, 2026}

\begin{abstract}
Precise predictions of cosmic-ray antinuclei fluxes, a prime
dark matter signature, are limited by the lack of
data constraining production rates of antinuclei. We mitigate this bottleneck with a fast, differentiable normalizing-flow surrogate for the femtoscopic source model (CECA), fit to 49 ALICE proton-proton correlation functions, and extrapolated via scaling laws to the low-multiplicity domain relevant for cosmic rays. The surrogate reproduces CECA with sub-percent emulation fidelity, yielding data-constrained source functions that remove the dominant uncertainty in coalescence-based antinuclei production rates.
The resulting uncertainties on the coalescence parameters $B_2$
and $B_3$ shrink from  factors of 10-100 and ${\sim}$1000 to the few percent and ten-percent level,  respectively, with an additional ${\sim}15\%$ wavefunction systematic for  $B_{2}$.
\end{abstract}

\maketitle 

Nearly a century after its initial detection in galaxy clusters~\cite{Zwicky:1933gu} and subsequent confirmation on galactic~\cite{Rubin:1970zza} and cosmological scales~\cite{Planck:2018vyg}, the nature of dark matter (DM) remains one of the most profound open questions in physics. The search for its conjectured particle constituents is pursued on three avenues~\cite{Cirelli:2024ssz}. Direct detection by scattering a DM particle off a nucleus~\cite{XENON:2025vwd, LZ:2025igz}, direct production in high-energy particle colliders such as the LHC~\cite{ATLAS:2023rvb, CMS:2024zqs},
and indirect searches via Standard Model particles produced in dark-matter annihilation or decays~\cite{Fermi-LAT:2025gei}.

Cosmic-ray antinuclei are a particularly promising candidate for the indirect search for dark matter. Theory predicts an enhancement of a factor $10$--$1000$ for the antinuclei flux at low kinetic energies when including DM decays~\cite{Donato:1999gy,Blum:2017qnn,Korsmeier:2017xzj}. This effect motivates dedicated antinuclei searches by experiments such as AMS-02~\cite{AMS02:2021pr} and GAPS~\cite{GAPS:2026rda}. While no antinuclei have been observed in cosmic rays to date, any future observation will require a precise prediction of the Standard Model secondary background before an unambiguous dark matter interpretation can be established. With the currently existing predictions~\cite{vonDoetinchem:2020vbj,Maurin:2025gsz}, any exotic signal can be masked by systematic model uncertainties.

Cosmic Rays and the interstellar medium are predominantly composed of protons/hydrogen ($\approx$ 90\%)~\cite{Ferriere:2001rg}, and about 70\% of all antinuclei are produced in collisions between cosmic ray protons with interstellar hydrogen at kinetic energies of approximately $300$~\GeV ($\sqrt{s}\approx24$~\GeV)~\cite{Serksnyte:2022onw}. Experimentally, the antinuclei production cross-sections at these energies are low, however, the programs at NA61/SHINE@SPS~\cite{NA61:2014lfx,Unger:2025ehn}, CBM@FAIR~\cite{Messchendorp:2951692}, and SMOG@LHC~\cite{Bursche:2649878,LHCb:2025udq}, will deliver first results in the energy range of interest in the next years.

In the absence of direct measurements, reliable predictions of the secondary antideuteron flux require an understanding of the microscopic formation mechanism. Currently existing predictions~\cite{vonDoetinchem:2020vbj,Maurin:2025gsz} introduce a model uncertainty of factors 10--100, due to an approximative treatment of nuclei formation. ALICE has shown~\cite{ALICE:2025byl} using {\ensuremath{\mathrm {\uppi\mbox{--}d}}\,\xspace} correlations, that the vast majority ($\approx88.9\pm6.3\%$) of (anti)nuclei are produced after the decay of short-lived resonances via binding processes. Such a process is commonly described by coalescence models~\cite{Bellini:2020cbj,Kachelriess:2020amp,mrowczynski_1987,Blum:2019suo,Mahlein:2023fmx,Scheibl:1998tk}, which describe the (anti)nuclei production as the phase-space overlap of the (anti)nucleus with the (anti)nucleons. The latter phase-space can be constrained using momentum correlation measurements (Femtoscopy) of proton-proton pairs~\cite{Fabbietti:2020bfg}. Combined with measured nucleon momentum distributions and a realistic nucleus wave function, coalescence reproduces the (anti)deuteron production cross-section in pp collisions at the LHC within $\approx$10\%~\cite{Mahlein:2023fmx}. Comprehensive studies by the ALICE collaboration have constrained the particle-emitting source at LHC energies and multiplicities~\cite{ALICE:2023sjd, ALICE:2026tqh}. However, the extrapolation of these results to the regime relevant to cosmic rays remains non-trivial. 

In this work, we present a data-driven framework to extrapolate the experimentally constrained emission source from TeV down to GeV using methods from machine learning. This approach enables improved predictions of antinuclei formation in the energy range relevant for cosmic-ray interactions and reduces the dominant uncertainty in secondary antinuclei flux calculations.

For (anti)deuterons, the coalescence probability is quantified by the parameter $B_2$, defined as the
ratio of the Lorentz-invariant (anti)deuteron spectrum to the squared (anti)nucleon spectrum
at equal momentum per nucleon~\cite{PhysRev.129.836, PhysRevC.21.1301}. More generally, the coalescence parameter for a nucleus of mass number $A$ is given by
\begin{align}
\label{eq:B2Experiment}
B_A = \left.
\frac{1}{2\pi p_{\mathrm{T},A}}
\frac{\mathrm{d}^2N_A}{\mathrm{d}y\,\mathrm{d}p_{\mathrm{T},A}}
\Big/
\left[
\frac{1}{2\pi p_{\mathrm{T}}}
\frac{\mathrm{d}^2N_p}{\mathrm{d}y\,\mathrm{d}p_{\mathrm{T}}}
\right]^A
\right|_{p_{\mathrm{T}} = p_{\mathrm{T},A}/A}.
\end{align}

To obtain the $B_2$ parameter and relate the (anti)nucleus creation probability to the final-state interaction, we introduce three assumptions: on-shell, equal-time, and smoothness~\cite{SATO1981153, mrowczynski_1987}. The on-shell and smoothness assumptions have been verified systematically, yielding sub-percent corrections at LHC energies~\cite{Smith:2026ocz}, while the equal-time assumption is a standard step in the derivation. This factorization reduces the convolution integral to a simple integral evaluated in the pair rest frame
\begin{equation}
    B_2 = \frac{4\pi}{3}
    \int \mathrm{d}^3\rs\; S_{\mathrm{pn}}(\rs)\;
    |\varphi_{\mathrm{d}}(\mathbf{\rs})|^2,
    \label{eq:B2}
\end{equation}
where \(\varphi_{\mathrm{d}}(\mathbf r)\) is the deuteron bound-state wave function~\cite{SATO1981153}.
The proton--neutron emission source $S_{\mathrm{pn}}(\rs)$ represents the probability distribution of finding a proton-neutron pair at a distance \rs after production and can be constrained via femtoscopy.
In femtoscopy one measures the two-particle correlation function \Cks,
as the ratio of correlated to uncorrelated pair yields as a function of the pair
rest-frame relative momentum \ks~\cite{ALICE:2018ysd}. It is related to the emission source through the
Koonin--Pratt equation~\cite{Koonin:1977fh, Pratt:1990zq}
\begin{equation}
    \Cks = \int \mathrm{d}^3\rs\; S(\rs)\; |\psi_{\ks}(\mathbf{\rs})|^2,
    \label{eq:KP}
\end{equation}
where $\psi_{\ks}(\mathbf{\rs})$ denotes the scattering wave function and encodes the strong and Coulomb interaction as well as quantum statistics of the pair.

Three aspects have so far prevented a data-constrained prediction of cosmic-ray antinuclei fluxes. First, cosmic-ray collisions occur at low particle multiplicity (\lowmult), a regime with no direct data coverage.
The emission source is governed by event multiplicity and the transverse mass of the pair (\mt) rather than
the collision energy~\cite{Lisa:2005dd, ALICE:2026tqh}. Since femtoscopic source functions are
precisely constrained across a wide multiplicity range at the LHC, with extensive
measurements from ALICE~\cite{ALICE:2020ibs,ALICE:2023sjd, ALICE:2025aur, ALICE:2026tqh}, ATLAS~\cite{ATLAS:2017shk, ATLAS:2022wvk},
CMS~\cite{CMS:2019fur}, and LHCb~\cite{LHCb:2017pnz}, the cosmic-ray regime becomes
accessible via extrapolation along empirically established \mt--multiplicity scaling
laws. 

Second, earlier coalescence calculations have employed source functions extracted from \SCpipi correlations as a proxy for
the \pN system~\cite{Scheibl:1998tk, Blum:2017qnn, Bellini:2018epz}. The obtained radii can exceed the values obtained from \pP correlations by
up to a factor of three, because of the contributions of long-lived resonances
($\omega$, $\eta$, $K_S^0$)~\cite{Wiedemann:1996ig, ALICE:2023sjd}.

Third, constraining the source
simultaneously across tens of correlation functions requires a model that can be evaluated at drastically lower computational cost. 

In this Letter, we mitigate all three aspects. We exploit \pP correlations as a substitute for the \pN source under isospin symmetry at matched
\mt and multiplicity and present a normalizing-flow surrogate trained on
CECA~\cite{Mihaylov:2023pyl} simulations. We simultaneously fit  49 ALICE \pP
correlation functions, double-differential in \mt and multiplicity. 
The normalizing flows technique is employed to map simple base distributions to complex, resonance-deformed emission geometries through a series of invertible bijective transformations. This allows us to perform fast, continuous, and fully differentiable density estimation of the source. Exploiting the
established $\mt$--multiplicity scaling, higher-statistics event classes anchor the
poorly populated low-multiplicity domain, delivering fully data-driven source functions
without additional extrapolation parameters. 
The framework is directly applicable to ongoing measurements at
SMOG@LHCb~\cite{LHCb:2025udq}, CBM@FAIR~\cite{Messchendorp:2951692}, and
NA61/SHINE~\cite{Shukla:2025imi, Kowalski:2916893}, providing the missing link between
collider femtoscopy and absolute cosmic-ray antinuclei flux predictions for space- and balloon-borne particle detectors such as
AMS-02~\cite{AMS02:2021pr} and GAPS~\cite{GAPS:2026rda}.

The 49 \pP correlation functions~\cite{ALICE:2020ibs, ALICE:2026tqh},
measured in proton–proton collisions at $\sqrt{s} = 13$ and 13.6~TeV,
span four multiplicity classes. These have been measured by the ALICE collaboration, three of them 
stemming from minimum-bias (MB) data~\cite{ALICE:2026tqh}, and one from a dedicated high-multiplicity (HM) triggered data set~\cite{ALICE:2020ibs}. 
Each of these four classes contain seven \mt ranges. The three MB classes include independently measured 
correlation functions for p–p and $\bar{\mathrm{p}}$–$\bar{\mathrm{p}}$, while the correlations within the HM class have a single 
correlation function per \mt bin, in which p–p$\oplus\bar{\mathrm{p}}$–$\bar{\mathrm{p}}$ are added together. 
Each correlation function $C(k^*)$ is related to the
underlying emission source \Srs and the p–p
scattering amplitude via the Koonin--Pratt equation~\eqref{eq:KP},
where $\psi_{\ks}(\mathbf{\rs})$ is the two-particle relative wave
function encoding the strong and Coulomb interaction as well as
antisymmetrization due to Fermi--Dirac statistics. The p--p
interaction governing $\psi_{\ks}(\mathbf{\rs})$ is known to high accuracy. High quality
phenomenological potentials such as Argonne $v_{18}$~\cite{Wiringa:1994wb} and
state-of-the-art chiral EFT potentials at N$^4$LO~\cite{Reinert:2017usi}
are available and yield consistent source estimates~\cite{ALICE:2026tqh},
making the inference of \Srs from Eq.~\eqref{eq:KP} robust. Correlation functions are computed via the CATS framework~\cite{Mihaylov:2018rva},
which numerically solves the Koonin--Pratt equation for a given source and
interaction. The fitting procedure follows Ref.~\cite{ALICE:2026tqh} exactly,
including feed-down and source dilution from $\Lambda$ decay products
via the $\lambda$ parameters. Crucially, the emission source in small systems scales
with event multiplicity but not with collision energy~\cite{Lisa:2005dd}, so the \TeV scale  ALICE data directly
constrain the source in the low-multiplicity regime relevant for
cosmic-ray applications, irrespective of the collision energy
at which antideuterons are produced in the ISM.
 
The emission source is parametrized with CECA (Common Emission in
CATS)~\cite{Mihaylov:2023pyl}, an effective source model, based on single particle properties, designed for the application in small
collision systems. CECA describes the source through three
parameters: $r_\mathrm{d}$, the core emission radius; $h$, a shape
parameter characterizing the hadronization surface, which adopts
a lens-like geometry and induces spatial--momentum correlations
encoding collective behaviour akin to radial flow; and $\tau$,
the effective system lifetime governing the displacement of
particles from the hadronization surface. The resonance contributions are
anchored by thermal-model calculations using
Thermal-FIST~\cite{Vovchenko:2019pjl}, with branching ratios and
hadronic states taken from the PDG~\cite{ParticleDataGroup:2024cfk}. The thermal
parameter estimation follows~\cite{Vovchenko:2019kes} and is consistent
with the ALICE source studies of~\cite{ALICE:2020ibs, ALICE:2023sjd}. 
The non-Gaussian tails are captured by the resonance contributions, while $(r_\mathrm{d}, h, \tau)$ encode the spatial–momentum correlations absent in simple Gaussian parameterizations.
CECA was previously used to fit Run~2 ALICE p--p data
and reproduces the empirically observed
\mt scaling of the source radius~\cite{ALICE:2020ibs, Mihaylov:2023pyl}. 
The microscopic interpretation of the scaling remains debated in the community,
though in large systems it is regarded as a hallmark of collective
radial flow~\cite{Lisa:2005dd}.

Evaluating the emission source doubly differentially in \mt and multiplicity via direct simulation is computationally intractable for simultaneous fitting of all 49 correlation functions. We replace the forward simulation with a conditional normalizing flow (NF)~\cite{pmlr-v37-rezende15}, using the implementation in JammyFlows~\cite{Glusenkamp:2020gtr}.  Normalizing flows provide a continuous and differentiable surrogate of the source distribution by learning an invertible transformation between a simple latent distribution and the complex, resonance-deformed emission geometry generated by CECA. The resulting surrogate describes the conditional source distribution
$S(\rs \mid r_\mathrm{d},h,\tau;\ks,\mt)$,
where $r_\mathrm{d}$, $h$, and $\tau$ are the three CECA source parameters, and the source is conditioned on the local pair kinematics as given by \ks and \mt. During inference the surrogate is marginalized over \ks~$\le$100~\MeVc, corresponding to the femtoscopically relevant region, delivering
a source profile conditioned solely on the CECA parameters and \mt. 

The
surrogate reduces the cost of a single source evaluation by more than two
orders of magnitude relative to direct simulation, making a grid scan over
the full CECA parameter space across all 49 correlation functions across four event classes computationally
feasible. While the surrogate is explicitly conditioned on local pair kinematics, the event-scale multiplicity dependence enters upstream. As detailed in the Supplemental Material~\ref{sm:AppendixML_MultParam}, the CECA parameters $(r_\mathrm{d}, h, \tau)$ evolve along a data-driven multiplicity axis mapping directly to the average charged-particle multiplicity ($\langle N_\mathrm{ch}\rangle$). This modular factorization allows the surrogate to supply the joint source distribution
$S(r^*;\mt,\langle N_\mathrm{ch}\rangle)$ at any arbitrary
$(\mt, \langle N_\mathrm{ch}\rangle)$ point on-the-fly during coalescence evaluation. 

Architecture details, training statistics, and closure tests
confirming sub-percent emulation fidelity are provided in the Supplemental
Material~\ref{sm:AppendixML_Details} and~\ref{sm:AppendixML_Closure}.

In order to evaluate $B_2$ and $B_3$ we use the ToMCCA model~\cite{Mahlein:2024pur, Mahlein:2025bla}, which provides the required Lorentz-invariant spectra of protons, deuterons, and $^3$He.
ToMCCA is a fast Monte Carlo event generator with a coalescence afterburner
designed to evaluate (anti)nuclei production via the Wigner function
formalism~\cite{Scheibl:1998tk}, taking the femtoscopic source as direct input. For clarity, we outline the framework focusing on the baseline two-body deuteron case. For each
proton--neutron pair with centre-of-mass momentum $\vec{P}$ and
relative momentum $\vec{q}$, the Lorentz-invariant deuteron yield
is given by~\cite{Mahlein:2024pur}
\begin{align}
    \frac{\mathrm{d}N_{\mathrm{d}}}{\mathrm{d}^{3}P} &=
    \frac{S_{\rm d}}{(2\pi)^{6}}
    \int \mathrm{d}^{3}r \int \mathrm{d}^{3}r_{\rm d} \int \mathrm{d}^{3}q\;
    \mathcal{D}(\vec{q},\vec{r}) \nonumber\\
    &\times W_{\rm np}\!\left(\tfrac{\vec{P}}{2}+\vec{q},
                  \tfrac{\vec{P}}{2}-\vec{q},
                  \vec{r},\vec{r}_{\rm d}\right),
    \label{eq:tomcca}
\end{align}
where $S_{\rm d}=3/8$ is the spin--isospin statistical factor,
$W_{\rm np}$ is the two-nucleon Wigner functions encoding
the phase-space distribution of the emitting source, and
$\mathcal{D}(\vec{q},\vec{r})$ is the internal Wigner density of the
deuteron,
\begin{equation}
    \mathcal{D}(\vec{q},\vec{r}) =
    \int \mathrm{d}^{3}\xi\; e^{-i\vec{q}\cdot\vec{\xi}}\;
    \varphi_{\rm d}\!\left(\vec{r}+\tfrac{\vec{\xi}}{2}\right)
    \varphi_{\rm d}^{*}\!\left(\vec{r}-\tfrac{\vec{\xi}}{2}\right),
    \label{eq:wigner_d}
\end{equation}
obtained from the deuteron wavefunction $\varphi_{\rm d}$. Simulation precision and convergence are reported in Supplemental Material~\ref{sm:AppendixTOMCCA_details}. Equation~\eqref{eq:tomcca} is evaluated
in the pair rest frame after applying the equal-time
approximation. Two wavefunction hypotheses are considered, 
N$^4$LO~\cite{Reinert:2017usi} and Argonne~$v_{18}$~\cite{Wiringa:1994wb} since both are
anchored to modern nucleon--nucleon scattering data and provide the
best description of ALICE deuteron spectra~\cite{Mahlein:2023fmx}. The NF surrogate supplies the spatial part of $W_{\mathrm{np}}$ at
arbitrary $(\mt, \langle N_\mathrm{ch}\rangle)$ on-the-fly, replacing
what would otherwise require a full event-generator simulation at
each phase-space point. While written here for $A=2$, this formalism scales analogously to multi-body systems and has been extended to treat 3-body coalescence~\cite{Mahlein:2025bla}, which we exploit directly to evaluate the $A=3$ ($^3\text{He}$ and $^3\overline{\text{He}}$) states presented in this work. In the latter, the two-body interaction is accounted for using Argonne~$v_{18}$ and off-shell consistent three-body forces are supplied by the Urbana~IX model~\cite{Pudliner:1995wk}. The extrapolation uncertainty on $B_A$ at \pt/$A$ $\lesssim$ 0.4 \GeVc
is estimated by comparing predictions obtained using the full model phase space, anchored to experimental data via mixed events, to those obtained by truncating the source to the lowest measured \mt range. The resulting spread is taken as the systematic uncertainty associated with the \mt extrapolation.

\begin{figure}[t]
    \centering
    \includegraphics[width=1.035\columnwidth]{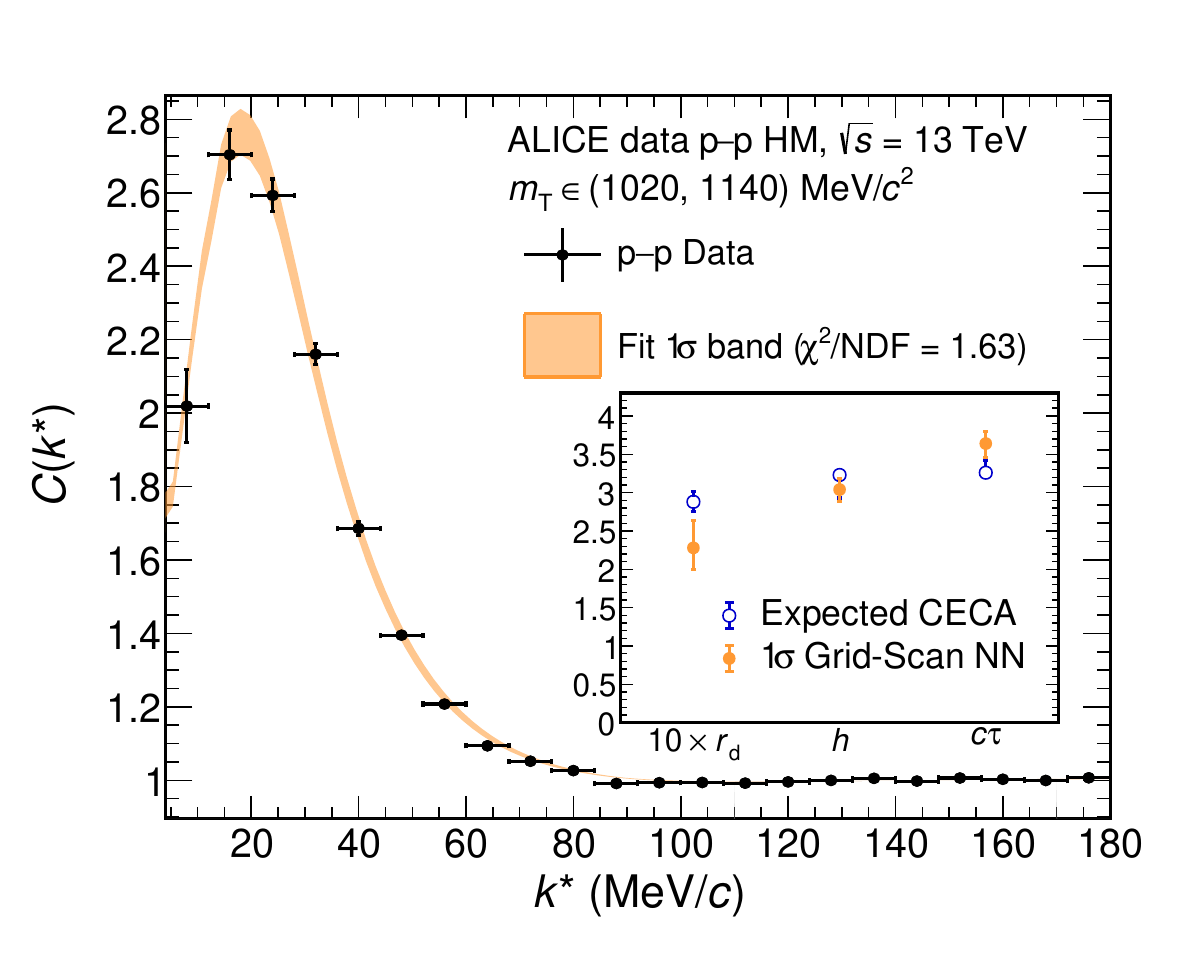}
    \caption{Representative p–p correlation function for the
    HM multiplicity class and selected \mt range. Data from ALICE~\cite{ALICE:2020ibs}
    (markers); NF surrogate fits (solid lines); uncertainty bands
    reflect the propagated parameter uncertainties from the grid
    scan (shaded). The inset shows the CECA parameters of the study~\cite{Mihaylov:2023pyl} compared to this work.}
    \label{fig:CFs}
\end{figure}

 
As a closure test and benchmarking of our method, we first repeat the CECA fit to the Run~2 ALICE p–p correlation functions~\cite{ALICE:2020ibs} using the NF surrogate in place of direct simulation. Figure~\ref{fig:CFs} shows an example of the fitted \pP correlation function. The obtained CECA source parameters are consistent, within uncertainties, with those reported in Ref.~\cite{Mihaylov:2023pyl}, validating both the surrogate emulation and the fitting pipeline. The uncertainties quoted in Ref.~\cite{Mihaylov:2023pyl} were obtained by simultaneously fitting \pP and \pL, yielding 13 correlations in total. In contrast, this work evaluates confidence regions using only the 7 available \pP correlations, avoiding potential systematics from the \pL interaction model.
 
Proceeding to the full dataset, the NF surrogate simultaneously describes all 49 correlation functions across the three MB multiplicity classes and seven \mt ranges ($\chi^2$/NDF = 1.39). The expected \mt and multiplicity dependence~\cite{ALICE:2026tqh} are extracted across the full kinematic range, confirming that CECA captures the relevant source geometry including resonance-driven non-Gaussian tails.
 The fitted source parameters, propagated through the eigenmode
decomposition described in the Supplemental Material~\ref{sm:AppendixML_MultParam}, yield
source functions at arbitrary multiplicity and \mt. 

\begin{figure}[t]
    \centering
    \includegraphics[width=\columnwidth]{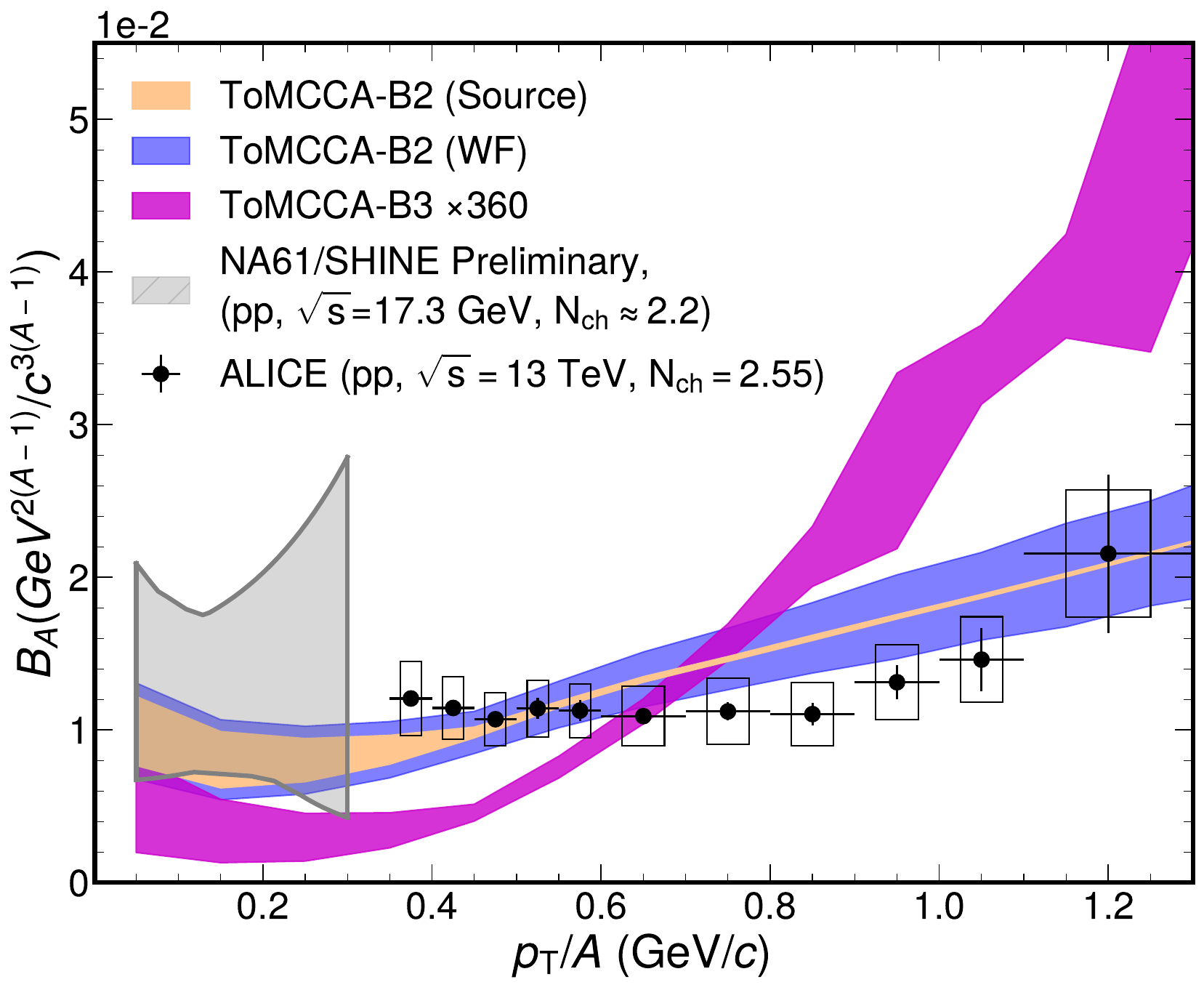}
    \caption{\pt-differential coalescence parameter $B_A$(\pt) for antideuterons ($B_2$, $A$=2) and antitritons/antihelium-3 ($B_3$, $A$=3). Shaded bands: ToMCCA predictions using NF-constrained CECA source functions (this work). For ($B_2$, the inner orange band shows the propagated source uncertainty and the outer blue band adds the wavefunction systematic (N4LO vs. Argonne $v$18) linearly. For $B_3$ (magenta), only the source uncertainty is shown, as the full calculation employs a single consistent 2+3 body force scheme (Argonne v18 + Urbana IX). The $B_3$ prediction is scaled by a factor of 30 for visual clarity. Filled markers: ALICE $B_2$ measurements~\cite{ALICE:2020foi}. The gray band represents the 1$\sigma$ confidence interval of $B_2$ estimated from preliminary NA61/SHINE data~\cite{Kowalski:2916893, Shukla:2025imi}, propagated via bootstrap resampling.}
    \label{fig:B2}
\end{figure}

Figure~\ref{fig:B2} shows the
resulting $p_T$-differential coalescence parameter $B_2(p_\mathrm{T}/A)$
compared to NA61/SHINE preliminary data~\cite{Kowalski:2916893, Shukla:2025imi} and the ALICE MB
measurement~\cite{ALICE:2020foi}, with which our prediction is consistent at the $0.84\sigma$ level. The NA61/SHINE results are preliminary and no $B_2$ is published, so it is constructed by fitting the proton~\cite{Kowalski:2916893} and deuteron~\cite{Shukla:2025imi} spectra with a \LT function and calculating Eq.~\ref{eq:B2Experiment}. The uncertainty estimation is done with a bootstrap procedure where points were varied within 1$\sigma$ during the fitting process. The full spread of the resulting $B_2$ is the band shown in Fig.~\ref{fig:B2}. The charged particle density interval of the NA61/SHINE data has been estimated to be $\langle \mathrm{d}N_\mathrm{ch}/\mathrm{d}\eta\rangle\approx2.2$ using the EPOS 3.117 event generator~\cite{Werner:2010aa} and requiring one charged particle at mid-rapidity $(|\eta|<1)$. The ALICE results are taken from the multiplicity class X, which covers the 70--100\% centrality interval, with $\langle \mathrm{d}N_\mathrm{ch}/\mathrm{d}\eta\rangle=2.55\pm0.04$. The ToMCCA predictions use $\langle\mathrm{d}N_\mathrm{ch}/\mathrm{d}\eta\rangle=2.2$, but the difference to the ALICE multiplicity is negligible compared to the extrapolation uncertainty.
The uncertainty of the predicted $B_2$ reaches up to $\sim$25\% for \pt below $0.4$~GeV/$c$ and decreases to at most $\sim$19\% for \pt above $0.4$~GeV/$c$. The latter kinematic region is covered directly by the available femtoscopic data.

Overall, the uncertainty on $B_2$ is reduced by more than an order of magnitude
relative to the factor of 10--100 variation achieved with previous approaches~\cite{vonDoetinchem:2020vbj, Maurin:2025gsz}.

The total uncertainty shown separately in Fig.~\ref{fig:B2} comprises two separate contributions: a source uncertainty (orange) and a wavefunction systematic (blue). The source uncertainty, propagated from the grid-scan performed with the NF surrogate, dominates at 
$p_T/A \lesssim 0.4$~GeV/$c$, reaching up to 23\%, and falls well below 3\% above this range. The wavefunction systematic of $\sim$15\%, arises from the spread between the N$^4$LO and Argonne $v_{18}$ deuteron wavefunction hypotheses. These two contributions are displayed as additive bands. The source uncertainty is statistical in nature and reducible with additional femtoscopic data, while the wavefunction systematic is an irreducible theoretical uncertainty of the present calculation.
 
The framework extends naturally to heavier antinuclei. For $B_3(\pt)$, it reduces the
uncertainty from a factor of $\sim$$1000$ to $\sim$$10\%$ in the measured kinematic range, degrading to $\sim$60\% below $p_T/A \lesssim 0.4$~GeV/$c$ where \mt extrapolation is required. This places the $\overline{^3\mathrm{He}}$ coalescence predictions on a quantitatively controlled footing for the first time via a data-constrained source. A dedicated assessment of the wavefunction/3 body-force systematic for $B_3$, analogous to the N$^4$LO–Argonne~$v_{18}$ comparison performed for $B_2$, is left to future work.

 
We have presented a femtoscopy-constrained pipeline for (anti)nuclei coalescence predictions, resolving the dominant source-function
uncertainty that has prevented precise predictions of the Standard Model secondary
background for cosmic-ray antinuclei. A conditional normalizing-flow
surrogate, trained on CECA simulations and simultaneously fit to
49 ALICE p–p correlation functions, delivers fully
data-constrained femtoscopic source functions across the
\mt--multiplicity plane at more than $100\times$ the speed of
direct simulation.

The framework is fully modular: the NF surrogate and the CECA multiplicity parametrization can be replaced by any improved source model, while the resulting $B_A$ predictions can be interfaced directly with cosmic-ray transport codes.

Propagated through the ToMCCA Wigner-function coalescence afterburner, these sources yield \pt-differential $B_2$ predictions with controlled uncertainties, reducing the dominant source uncertainty by more than an order of magnitude relative to previous approaches. The same framework extends naturally to $B_3$, reducing the corresponding uncertainty by nearly two orders of magnitude and providing the first quantitatively controlled, data-constrained predictions based on femtoscopic source information.

These results provide realistic, quantitatively controlled inputs for cosmic-ray antinuclei studies and are directly applicable to ongoing and planned measurements at SMOG@LHCb, CBM@FAIR, and NA61/SHINE. The resulting $B_A$ predictions provide a controlled basis for future antideuteron flux predictions for AMS-02 and GAPS. The differentiable structure of the NF surrogate
is a key ingredient toward a fully end-to-end differentiable inference
pipeline. Completing this vision requires differentiable
implementations of the correlation-function evaluator, the ToMCCA
event generator, and cosmic-ray propagation codes such as DRAGON~\cite{Evoli:2008dv} or
GALPROP~\cite{Strong:1998pw}.

\begin{acknowledgments}
The authors gratefully acknowledge funding support from the “Neutrinos and Dark Matter in Astro- and Particle Physics” (SFB 1258) (Grant No.~283604770). We thank organizers and participants of the JENAA workshop on nuclear physics at CERN in August 2024 for enabling insightful discussions regarding current status of coalesence estimates within the context of cosmic rays. DM acknowledges that this work was partially supported by the Bulgarian National Roadmap for Research Infrastructures - Object CERN.
\end{acknowledgments}

\bibliography{bibliography}

\clearpage

\appendix

\begin{center}
    {\large\textbf{Supplemental Material}}\\[0.5em]
    {\textit{Femtoscopy-Constrained Light Antinuclei Coalescence: Bridging Colliders and Cosmic Rays with Neural-Flow Surrogates}}
\end{center}

\vspace{0.5em}

\section{CECA Multiplicity Parametrization via Likelihood Eigenmode Decomposition}
\label{sm:AppendixML_MultParam}
 
While the CECA model~\cite{Mihaylov:2023pyl} carries no native exposed multiplicity dependence, with the emission source being fully characterized by the structural parameters $(r_{\rm{d}}, h, \tau)$, the experimental data inherently spans a wide multiplicity range. To address this, we introduce a data-driven multiplicity axis by exploiting the topological structure of the joint likelihood landscape.
 
For each of the four multiplicity classes (three MB, one HM) we
scan the CECA parameter space jointly over all $N_{\mt} = 7$
transverse-mass bins. For the MB classes p–p and
$\bar{\mathrm{p}}$--$\bar{\mathrm{p}}$ correlation functions are
treated separately~\cite{ALICE:2026tqh}; for the HM class the two
backgrounds are indistinguishable and p–p$\oplus\bar{\mathrm{p}}$–$\bar{\mathrm{p}}$ are added together,
giving 49 correlation functions in total. The joint log-likelihood
at each grid point for each multiplicity class $(\mathcal{M})$ is
\begin{equation}
    \ln \mathcal{L}^{\mathcal{M}}(r_{\rm{d}}, h, \tau) =
    \sum_{i=1}^{N_{\mt}} \sum_{s \in \mathcal{S}}
    \ln \mathcal{L}_{i,s}^{\mathcal{M}}((r_{\rm{d}}, h, \tau \mid \mt^{(i)}),
    \label{eq:joint_likelihood}
\end{equation}
where $\mathcal{S} = \{\mathrm{pp},\bar{\mathrm{p}}\bar{\mathrm{p}}\}$
for MB and $\mathcal{S} = \{\mathrm{pp}\}$ for HM.
 
Inspection of the landscape shows that $r_{\rm{d}}$ is only weakly
constrained (the likelihood is tiled along $r_{\rm{d}}$), while the
physically relevant information resides in the correlated
$(h,\tau)$ subspace. We marginalize over $r_{\rm{d}}$ within its
$1\sigma$ credible region. In the resulting $(h,\tau)$ plane
the likelihood contours form elongated ellipses whose principal
axes define two eigenmodes: $\eta$, along the valley (soft
direction), and $\xi$,
across the valley (well-constrained direction), encoding the multiplicity evolution. A global rotation
by angle $\theta$ maps $(h,\tau) \to (\eta,\xi)$; consistency of
$\theta$ across all multiplicity class fits is verified before applying the
rotation.
 
\begin{figure}[h]
    \centering
    \begin{minipage}{0.49\textwidth}
        \centering
        \includegraphics[width=\textwidth]{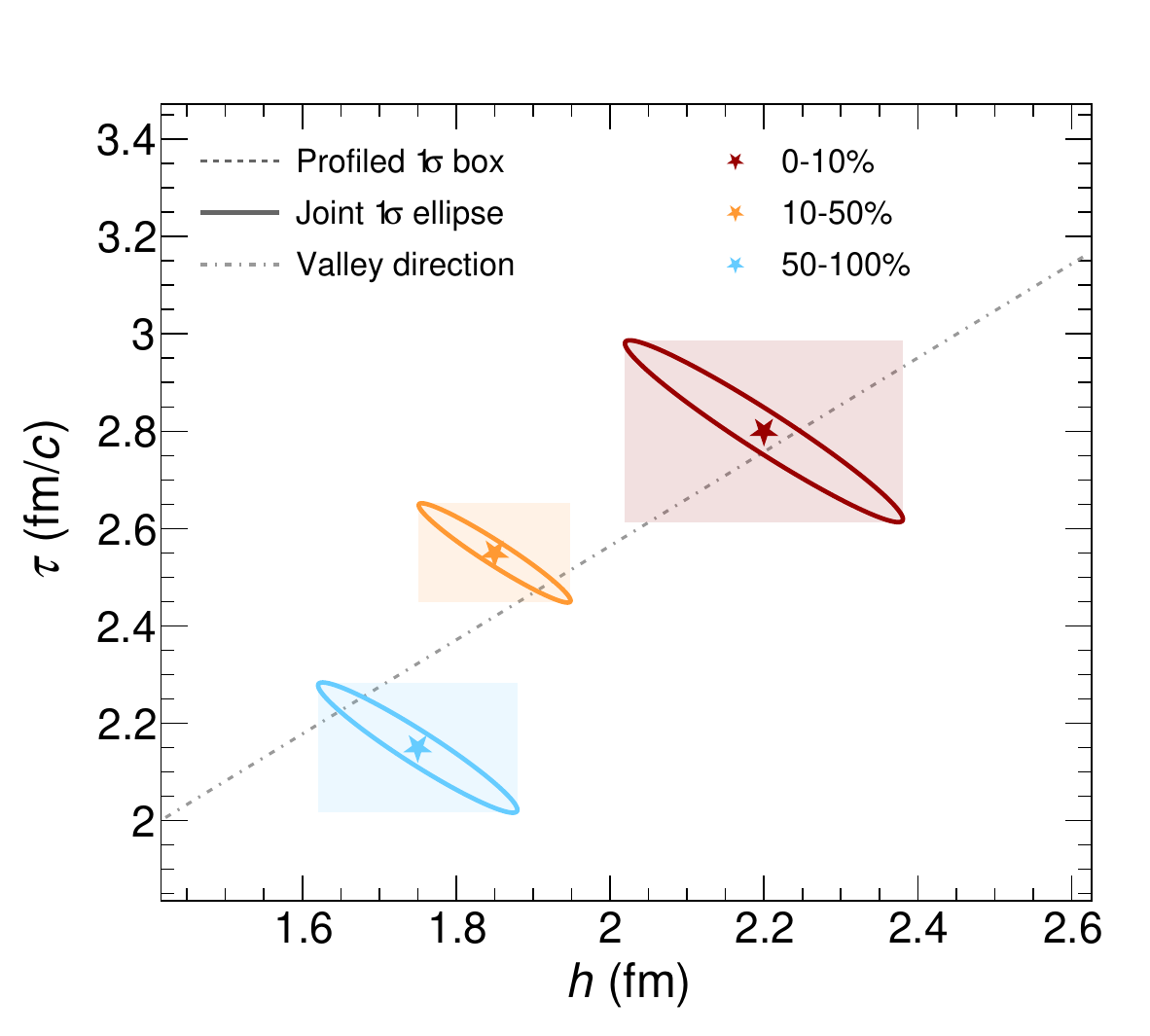}
    \end{minipage}
    \hfill
    \begin{minipage}{0.49\textwidth}
        \centering
        \includegraphics[width=\textwidth]{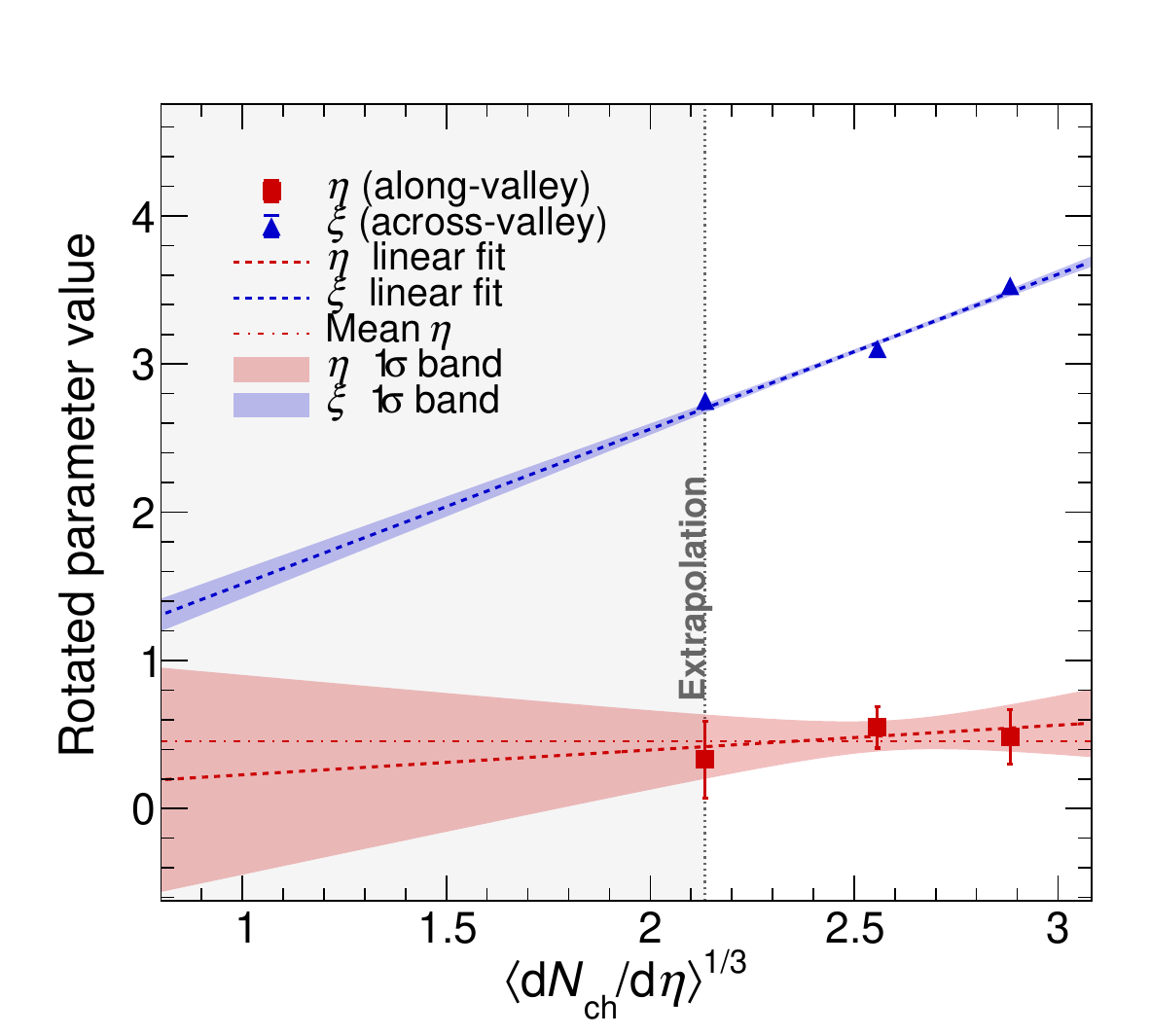}
    \end{minipage}
    \caption{Likelihood eigenmode decomposition and multiplicity scaling. \textit{Upper:} Joint $1\sigma$ likelihood ellipses and profiled 1D $1\sigma$ error boxes in the unrotated $(h, \tau)$ parameter plane for the three MB classes. The dashed line explicitly tracks the common diagonal valley direction. \textit{Lower:} Extracted data points and linear fits for the rotated parameters as a function of $\langle\mathrm{d}N_{\mathrm{ch}}/\mathrm{d}\eta\rangle^{1/3}$. The across-valley parameter $\xi$ carries the steep multiplicity-driven scaling, while the along-valley parameter $\eta$ tracks the flat residual along the poorly constrained direction.}
    \label{fig:pca}
\end{figure}

In $(\eta,\xi)$ coordinates the multiplicity dependence is well
described by a linear model. A global linear fit in each
eigenmode, with full likelihood covariance propagated from the
grid scans, anchors the scaling to the three MB classes and
extrapolates to \lowmult without additional free parameters. The
HM class is excluded from this fit, it occupies a distinct
phase-space and its $\langle N_\mathrm{ch}\rangle$ is not
precisely known, and hence only used for the consistency check of our procedure as reported in the manuscript.
Figure~\ref{fig:pca} shows the likelihood contours in the
$(\eta,\xi)$ plane and the linear scaling with
$\langle N_\mathrm{ch}\rangle$.
 
\section{Normalizing-Flow Architecture and Training}
\label{sm:AppendixML_Details} 

The conditional normalizing flow (NF) learns the map
\begin{equation}
    \mathbf{z}_0 \xleftrightarrow{\mathrm{NF}}
    \mathbf{x}\equiv\rs \mid (r_\mathrm{d},h,\tau,\ks,\mt),
    \label{eq:nf_map}
\end{equation}
where $\mathbf{z}_0$ denotes latent variables sampled from a simple base
distribution $p_0(\mathbf{z}_0)$, chosen as a multivariate Gaussian, and
$(r_{\rm d},h,\tau)$ are the three CECA source parameters. Here, $\ks$ is
the pair rest-frame momentum, $\mt$ is the pair transverse mass, and
$\mathbf{x}$ represents samples drawn from the resulting femtoscopic source
distribution $S(\rs\mid r_\mathrm{d},h,\tau;\ks,\mt)$.
The network transforms a tractable base distribution $p_0(\mathbf{z})$
into the target source density through a composition of $K$ invertible
transformations $\mathcal{F}_\theta = f_K \circ \cdots \circ f_1$,
with the exact analytically tractable log-likelihood
\begin{equation}
    \ln p(\mathbf{x}) = \ln p_0(\mathbf{z}) -
    \sum_{k=1}^{K} \ln \left|\det
    \frac{\partial f_k}{\partial \mathbf{z}_{k-1}}\right|,
    \label{eq:loglik}
\end{equation}
where $\mathbf{z}_0 \sim p_0$ and $\mathbf{x} = \mathcal{F}_\theta(\mathbf{z}_0)$.

\paragraph{Architecture.}
The network consists of one affine coupling
layer followed by two Gaussian mixture layers,
totaling 9\,024 trainable parameters. In the affine coupling
layer the input is split into two partitions; one partition passes
through unchanged while the other undergoes an element-wise affine
transformation whose scale and shift are predicted by a small
sub-network conditioned on the first partition. The two subsequent
Gaussian mixture layers model the residual non-Gaussianity in the
transformed space, capturing in particular the resonance-induced
non-Gaussian tails of the CECA source distributions that a purely
Gaussian flow would fail to reproduce. The conditioning of $(r_{\rm{d}}, h, \tau)$  on
$(\ks, \mt)$ is implemented by concatenating these
variables to the input of each sub-network at every layer.

\paragraph{Training data.}
Training samples are generated from $2300$ CECA configurations
drawn via Latin-hypercube sampling of the three-dimensional
$(r_{\rm{d}}, h, \tau)$ parameter space, each evaluated at $80\,000$
phase-space points $(\ks, \mt)$, yielding $1.84 \times 10^8$
training samples in total. An independent validation set of $700$
held-out CECA configurations is reserved exclusively for early
stopping and hyperparameter selection; these configurations are
never seen during training. The network is trained by minimizing
the exact negative log-likelihood using the AdamW Schedule-Free optimizer with an initial learning rate of 0.0025 and weight decay 0.05. Training converges in
approximately $20$ epochs on a single GPU,
after which the surrogate is frozen and used as a static emulator.
The total training cost is a one-time investment; subsequent
evaluations are effectively instantaneous, yielding a speed-up of
more than two orders of magnitude over direct CECA simulation.
Table~\ref{tab:training} summarises the key architectural and
training hyperparameters.

\begin{table}[h]
    \centering
    \caption{Summary of normalizing-flow architecture, training setup, and performance.}
    \label{tab:training}
    \begin{ruledtabular}
    \begin{tabular}{lc}
        Hyperparameter & Value \\
        \hline
        Coupling layers         & 1 affine $+$ 2 Gaussian mixture \\
        Trainable parameters    & 9\,024 \\
        Training configurations & 2\,300 \\
        Samples per config.     & 80\,000 \\
        Total training samples  & $1.84\times10^{8}$ \\
        Validation configs.     & 700 \\
        Optimizer               &  AdamW Schedule-Free \\
        Speed-up vs.\ CECA      & $>100\times$ \\
    \end{tabular}
    \end{ruledtabular}
\end{table}

\section{Closure Tests and Emulation Fidelity}
\label{sm:AppendixML_Closure} 

As a blind Monte Carlo closure, we select one arbitrary CECA configuration from the prior, produce synthetic correlation functions with realistic uncertainties drawn from the Run~2 measurement errors, and fit with the NF surrogate without knowledge of the ground-truth parameters. The recovered parameters agree with the injected values to within $1\sigma$, and the uncertainty interval exhibits correct frequentist coverage, confirming that the surrogate-based likelihood yields well-calibrated parameter estimates.

The physics closure is described in the main text,  repeating the Run~2 ALICE \pP fit of Ref.~\cite{Mihaylov:2023pyl} with the NF surrogate in place of direct simulation recovers the published source parameters and \mt scaling to within uncertainties.

\section{ToMCCA Coalescence Afterburner: Implementation Details}
\label{sm:AppendixTOMCCA_details}
 
The integrals in Eqs.~(4) and~(5) of the main text are evaluated
in the pair rest frame after applying the equal-time
approximations~\cite{Mahlein:2024pur, Mahlein:2025bla}. The internal Wigner density
$\mathcal{D}(\vec{q},\vec{r})$ is computed on a two-dimensional fully angular averaged
momentum--position grid with step sizes $\Delta q = 3.33$~MeV/$c$
and $\Delta r = 0.067$~fm, verified to be converged to better than
0.1\textperthousand\, using the \texttt{SciPi.optimize.curve\_fit} adaptive integrator~\cite{2020NaMet..17..261V}. Two hypotheses for the deuteron wavefunction are evaluated: chiral EFT@N$^4$LO~\cite{Reinert:2017usi} and Argonne
$v_{18}$~\cite{Wiringa:1994wb}. Both interaction models provide an excellent description of ALICE deuteron high-multiplicity
spectra~\cite{Mahlein:2023fmx} and are anchored to modern
proton--neutron scattering data; the spread across wavefunction
choices contributes a systematic uncertainty that evolves from sub-dominant at low $\pt/A$ to the dominant uncertainty at high $\pt/A$, reaching at most $\sim 15$\%. The normalizing flow surrogate supplies the spatial part of $W_{\mathrm{np}}$ at arbitrary
$(\mt, \langle N_\mathrm{ch}\rangle)$ on-the-fly at each
integration point, replacing what would otherwise require a full
CECA simulation call. The full pipeline from source evaluation to
$B_2$ output takes $\sim 0.2$~ms per $(\mt,\,\langle
N_\mathrm{ch}\rangle)$ point on a standard CPU core, compared to
$\sim 1200$~s for a direct CECA simulation,
confirming the $>100\times$ speed-up quoted in the main text.
The resulting $B_2(\mt)$ values are stored as a function of \mt
and $\langle N_\mathrm{ch}\rangle$ and may be ingested directly into
cosmic-ray propagation codes.  

\end{document}